# Real-time operating mode with DSSSD detector to search for short correlation ER-alpha chains

Yu.S.Tsyganov, A.N.Polyakov

FLNR, JINR, 141980 Dubna, Russian Federation

tyura@sungns.jinr.ru

**Abstract**

*Real-time PC based algorithm is developed for DSSSD detector. Complete fusion nuclear reaction $^{nat}Yt+^{48}Ca \rightarrow ^{217}Th$ is used to test this algorithm at $^{48}Ca$ beam. Example of successful application of a former algorithm for resistive strip PIPS detector in $^{249}Bk+^{48}Ca$ nuclear reaction is presented too. Case of alpha-alpha correlations is also under brief consideration.*

1. **Introduction**

The existence of superheavy elements (SHE) was predicted in the late 1960s as one of the first outcomes of the macroscopic-microscopic theory of atomic nucleus. Modern theoretical approaches confirm this concept. To date, nuclei associated with the "island of stability" can be accessed preferentially in $^{48}Ca$-induced complete fusion nuclear reactions with actinide targets. Successful use of these reactions was pioneered employing the Dubna Gas-Filled Recoil Separator (DGFRS) at the Flerov Laboratory of Nuclear Reactions (FLNR) in Dubna, Russia]. In the last two decades intense research in SHE synthesis has taken place and lead to significant progress in methods of detecting rare alpha decays. Method of "active correlations" used to provide a deep suppression of background products is one of them. Significant progress in the detection technique was achieved with application of DSSSD detectors. In particular, DSSSD detector was applied in the experiment with the reaction $^{249}Bk+^{48}Ca \rightarrow 117+3,4n$ aimed at the synthesis of Z=117 element. Note that applying the method of "active correlations" with DSSSD detector is even more effective compared with the case of resistive strip PIPS detector. On the other hand, some specific effects take place and possible sharing registered signal between two neighbor strips is one of them. The aim of this paper is to present development of method of "active correlations" for application with DSSSD detector. The Dubna Gas Filled Recoil Separator (DGFRS) is one of the most effective facilities in use for the synthesis of super heavy elements (SHE) [1]. Using this facility it has been possible to obtain 49 new super heavy nuclides. The PC based detection system allows storing event by event data from complete fusion nuclear reactions aimed to the study of rare decays of SHE [2]. The parameter monitoring and protection system of the DGFRS [3] has been designed in order to provide for the operation safety in the long term experiments with high intensity heavy ion beams and highly active actinide targets, as well as to provide for monitoring of the experimental parameters associated with the DGFRS, its detection system and the U-400 FLNR cyclotron.

The DGFRS was put into operation in 1989. It separates products under investigation from target and beam-like charge particles due to the difference in their magnetic rigidity value. It constructed according to D-Q-Q design (dipole magnet and quadrupole doublet). Since then many significant improvements and numerous model experiments were accomplished to develop the separator into a facility for heavy element research. Owing to its underlying principle, the separator shows excellent qualities for fusion-evaporation reactions induced by $^{48}$Ca. Special emphasis was laid on the possibility of applying very intense beams of heavy projectiles to strongly radioactive and rather exotic target species like $^{242}$Pu, 249Bk, $^{248}$Cm or $^{249}$Cf.

The separator consists of a $23^0$ dipole magnet and quadrupole doublet. The angular acceptance of the facility is +/-30 *mrad* vertically and +/- 60 *mrad* horizontally. The dispersion in magnetic rigidity is 8.5 *mm* per % *Bρ* deviation. The target is rotated usually with a speed 1680 *rev/min*.

2. **Method of "active correlations"**

From the viewpoint of detection system the experiment on synthesis and study of the properties of superheavy nuclei is one of the most difficult tasks. In fact, these experiments can be considered extreme in many cases:
- extremely low formation cross sections of the products under investigation,
- extremely high heavy ion beam intensities,
- high radioactivity of actinide targets, which are used in the experiments aimed at the synthesis of superheavy nuclei,
- extremely long duration of the experiment,
- extremely low yield of the products under investigation,
- very high required sensitivity of the detection system and
- radical suppression of the background products ( method of "active correlations"),
- high reliability level of monitoring system,
- high quality visualization system for spectroscopy on-line data flow.

The $^{48}$Ca ion-beam accelerated by U-400 cyclotron of Flerov Laboratory of Nuclear Reactions (FLNR). The typical beam intensity at the target was 1-1.3 pμA. The evaporation residues (EVRs) recoiling from the target were separated in flight from $^{48}$Ca beam ions, scattered particles and transfer reaction products by the DGFRS. EVRs passed through a time-of-flight system and were implanted to PIPS 32 strip position sensitive detector. To provide a deep suppression of beam associated background product Pc based real-time algorithm has been designed. It based on a simple idea. Namely, it consists in searching the time-energy-position EVR-alpha links in a real-time mode, and using the discrete representation of the resistive layer of position sensitive PIPS detector separately for signals like recoils and alpha particles. Thus, the PIPS detector is represented in the RAM of a PC in the form of two EVR matrixes: one for "top" position signal, another one for "bottom" position signal. The first index of each matrix is defined as a strip number, whereas the second index is defined by the event vertical position. The incoming EVR elapsed time value is written to those matrix elements. In the case of "alpha" signal detection a comparison with both EVR matrix elements is made, involving neighbor elements (+/- 3). If the minimum time is less or equal to the setting time, the detection system turns on the beam chopper which deflects the heavy ion beam in the injection line of the cyclotron for a few minutes. At the next step the PC code ignores the vertical position sensitivity of the alpha particle signals during the beam-off interval. If such decay takes place in the same strip that generated the pause, the duration of the beam-off interval is prolonged up to tens of minutes. In the Fig.1 an example of application of such method for $^{249}$Bk+$^{48}$Ca experiment is shown. Energies, time and positions of the decay signals are shown in the figure. Shadows denote that corresponding signals (strip #8) are detected in the beam-off interval.

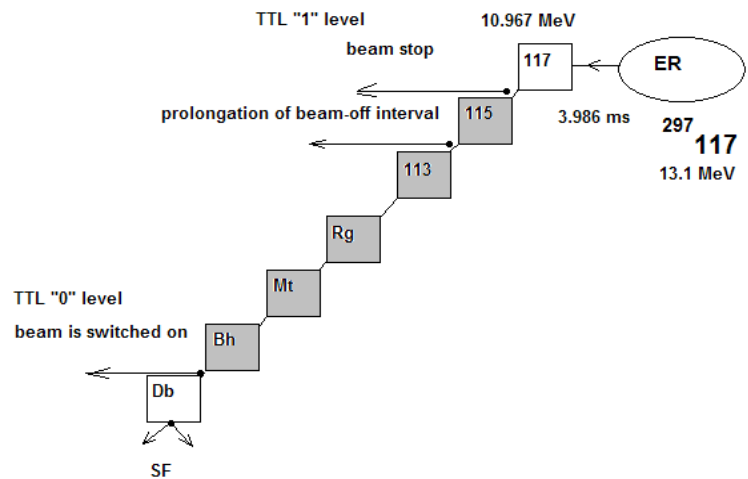

**Fig.1** The example of the "active correlations" method application for position sensitive 32 strip PIPS detector. The last signal in the chain is $^{270}$Db spontaneous fission. Target irradiation by $^{48}$Ca beam was stopped by ER ($^{297}$117)-alpha ($^{294}$117) detected correlation sequence. Values of decay energy and elapsed time are shown.

## 3. DSSSD detector application. Test reaction $^{nat}Yt + {}^{48}Ca \rightarrow {}^{217}Th+3n$.

As to the specificity of applying DSSSD detector (48 front side strips and 128 back side strips) and development of a corresponding real-time algorithm, one should keep in mind the following:

- detector's structure corresponds to matrix of the given dimensions which, in first approximation, can be used as the matrix of recoil nuclei; its elements are filled in by value of the current time taken from CAMAC hardware upon receiving the corresponding events;

- due to presence of P+ isolating layer between two neighbor strips on the ohmic side of the detector (48 front strips) the edge effects are negligible;

- on the contrary, for the 128 back strips (p-n junction) the effect of charge sharing between neighboring strips can be up to some 17% in the geometry close to $2\pi$. Certainly, this effect should be taken into account when developing and applying the algorithm of search of potential ER-α correlation. In addition to Table 1 indicating main electronic modules, below is presented the 14-word (16 bit each) event specification as C++ code fragment.

Below, in the Fig.2,3 two schematics of the process to operate together with the DSSSD detector are shown. Dead times are indicated in the figure b). Note, that the resulting dead time value is equal to 100 µs. Signals from front strip and back strip, as well as signals TOF, dE1 and dE2 from gaseous detector. To minimize the back neighbor strip edge effects role the same elapsed time value is written not only to matrix element "i,j", but "i, j+/-1" too. Here "I" is a first (front) EVR matrix index.

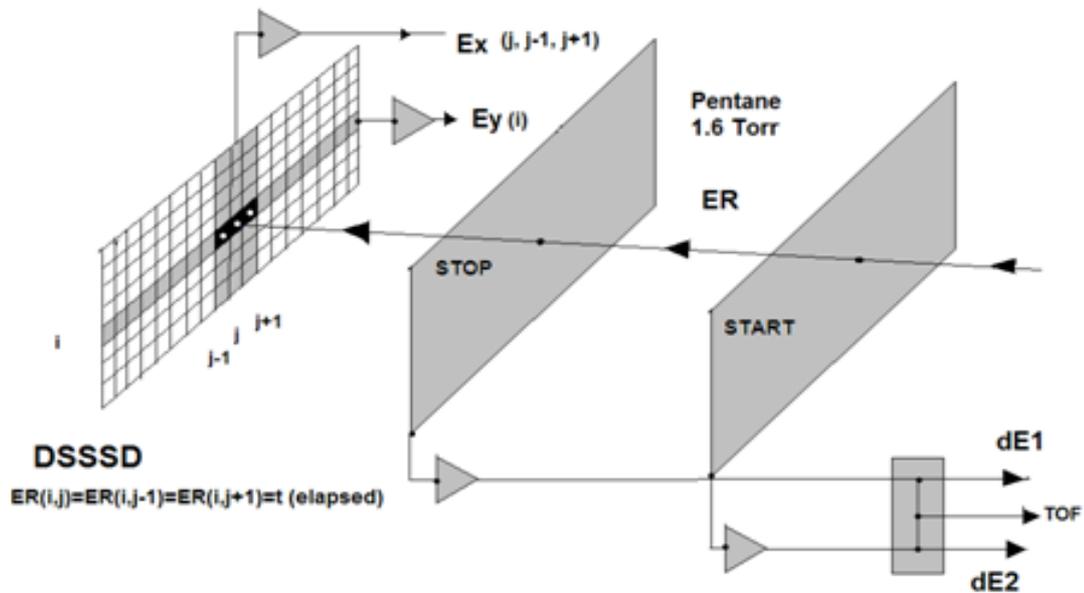

**Fig.2** Schematic of the ER matrix element value formation. DSSSD focal plane detector and START and STOP low pressure gaseous chambers are shown.

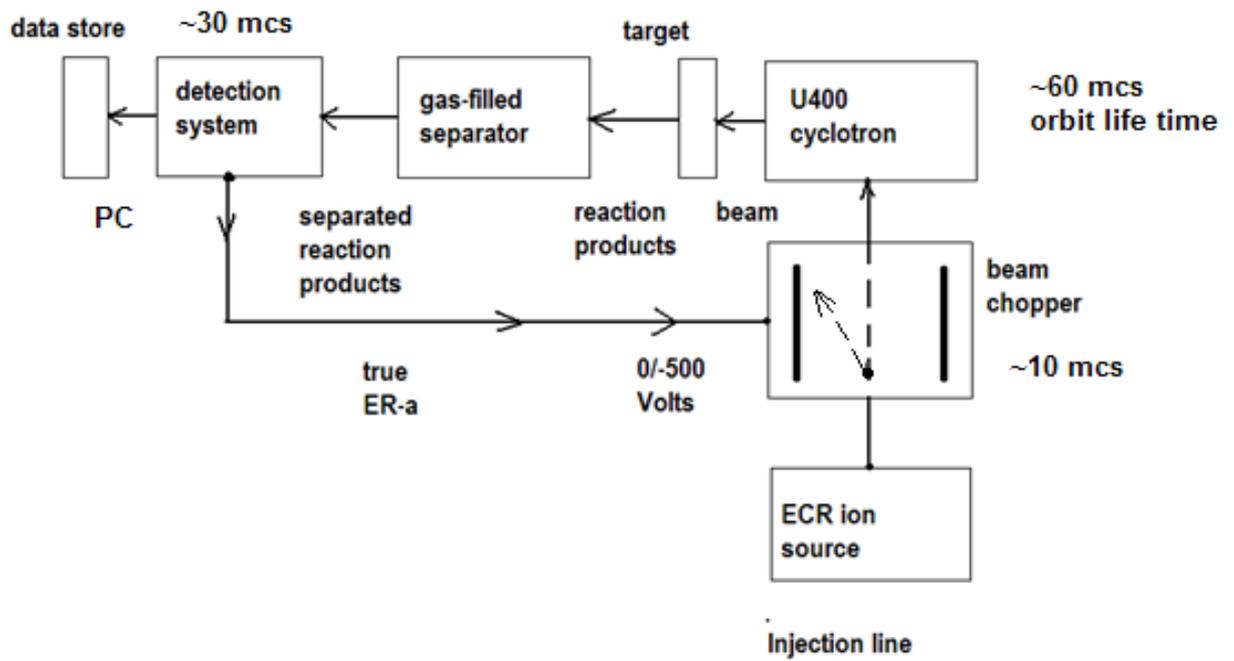

**Fig. 3.** The schematics of the beam chopping process, when ER-α correlation chain is detected. Dead times are indicated. Time of flight of the ER trough the DGFRS is about 1-2 μs.

The beam test of the briefly described method was performed in the $^{nat}Yt+^{48}Ca \rightarrow ^{217}Th+3n$ reaction. Below, extracted for 0.47 ms time interval of $^{217}Th$ EVR spectrum is shown in the Fig.3. The presented spectrum shape is typical for heavy recoils detected with silicon radiation detector.

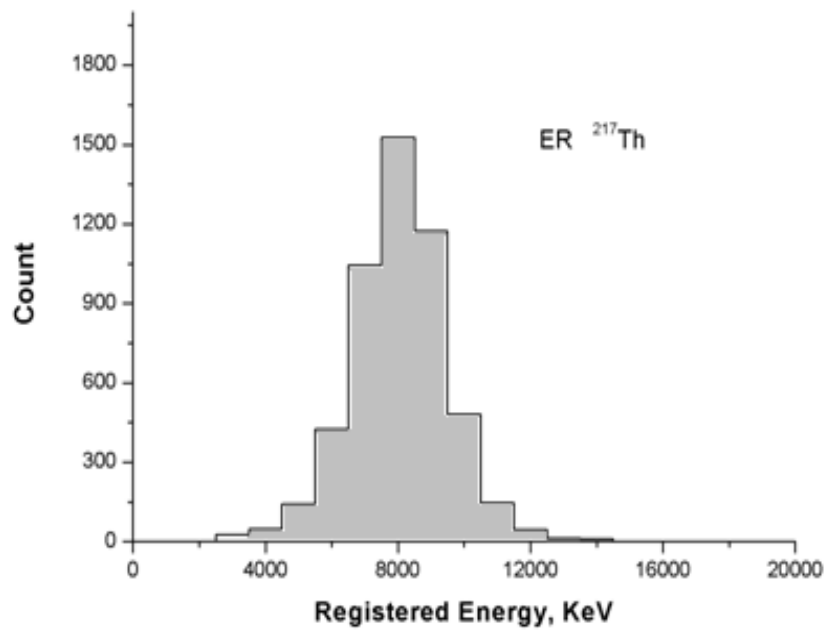

**Fig.4** Spectrum of $^{217}Th$ recoil registered energy

## 4. Case of alpha-alpha correlations detection if efficiency of ER detection is not close to 100%.

Let us consider a case of a few subsequent alpha decay chains when efficiency of ER detection is not close to 100% like it is considered above. The corresponding decay picture is shown in the Fig.4. If one considers 2D picture, Fig.5 except for Fig.4, connects all вершины ((n·(n-1)/2 links in total) by oriented lines and places $α^k_{ij}$ matrixes onto the graph вершину. Here k is the number of the detected signal which can be attributed to alpha-decay of SHE. It is possible to compose for each alpha particle signal candidate the relationship like $\Delta t_{i,j}^{k,k+n}=\min \{α_{i,j}^{k} - α_{i,j+m}^{k+n}\}|_{m=0,1,-1}$. Hence, if at the given time moment this parameter is less or equal than setting parameter $t_{kn}$, then system can generate beam stop for a short time.

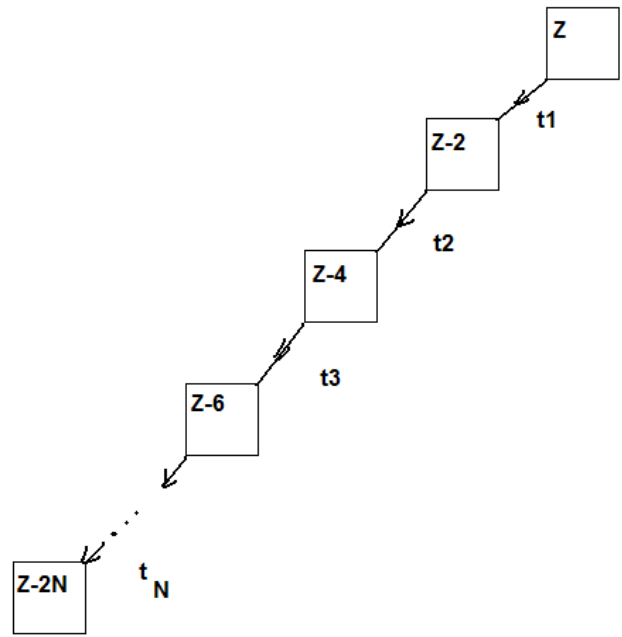

**Fig.5** Alpha decay chains 1…N for Z to Z-2·N.

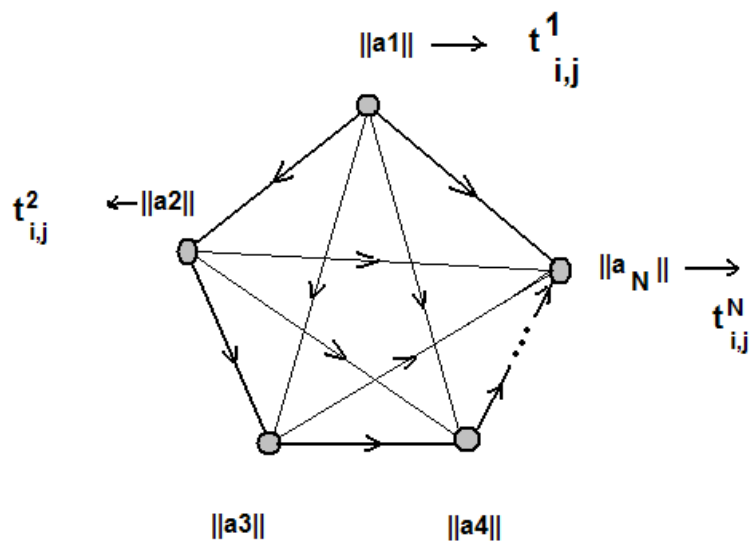

**Fig.6** Schematics of the algorithm for $\frac{N·(N-1)}{2}$ α-α correlated chains.

If, according to the requirements of an experiment beam-stop after; missing $n$ alpha particles is available and $\psi$ is an efficiency to detect alpha particle by a focal plane detector, then one can consider the value of $P_n = \psi \cdot \sum_{i=0}^{n}(1-\psi)^i$ as a probability to generate the mentioned beam-stop signal. In the real experiments parameter of $\psi$ is close to 0.5, although if one takes into account detection by not only by focal plane detector, nut by side detector too, then it may be as about 0.7-0.85 depending on the energy threshold of the detection system. And of course, it is easy to include the ER signal into the above mentioned process consideration with the parameter of the detection efficiency $\psi_{ER} \approx 1$.

Another interesting case from the viewpoint of long term experiment application is a combined process. The block diagram of the process is presented schematically in the Fig.7. Each event specified as 14 Word of 16 bit is in fact considered as incoming system for both single (ER signal) and correlation (ER-alpha correlated sequence detection) algorithms. The response for the incoming event in the case of the given signal (pair of signals) corresponds to the true recoil signal (ER) or true energy-time-position correlation sequence (ER-alpha) the C++ Builder 6 written code generates "pause" signal for a short times $t_1$ and $t_2$, respectively. Note, than the value of $t_1$ is much less than $t_2$ in order to minimize a loss in a full efficiency of the experiment due to a break point in the target irradiation process.

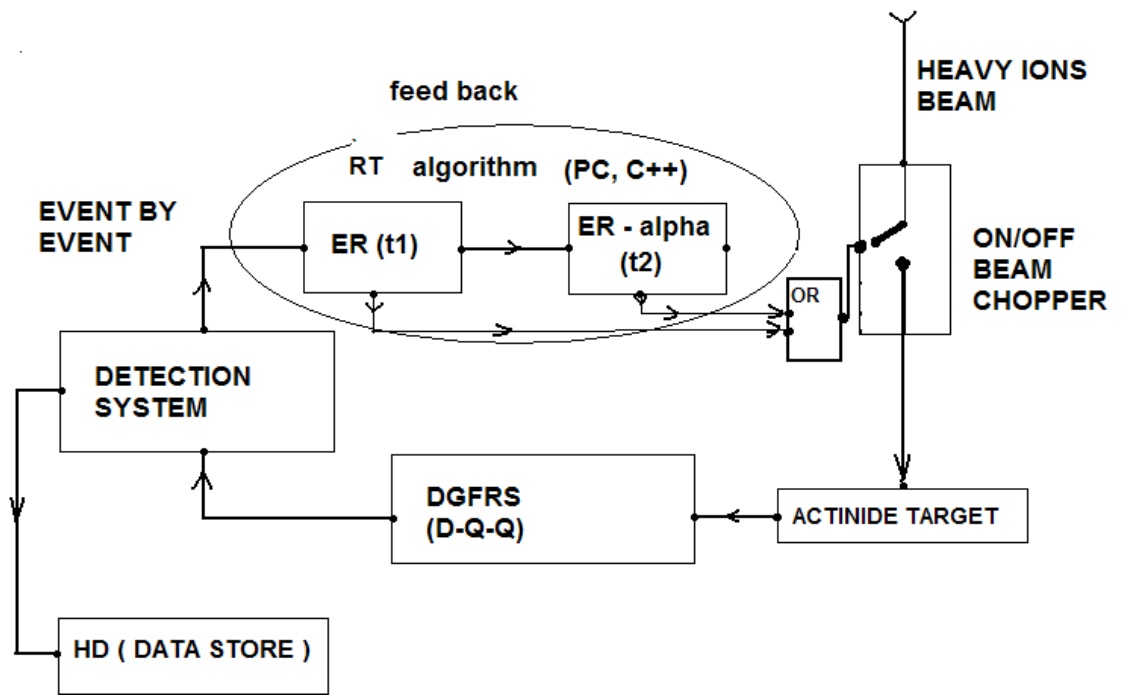

**Fig.7** Block diagram of the combined process (schematically); $t_1 \ll t_2$.

It is very easy to estimate roughly losses in the whole irradiation process.

With the presented schematics in the Fig.**7** one can state to a first approximation that:

$$t_d \approx T_{exp} \cdot (\nu_{ER} t_1 + \nu_{ER} \nu_\alpha \tau_{ER-\alpha} t_2).$$

Here:

$t_d$ – dead time for one DSSSD element (pixel) representation, $\nu_{ER,\alpha}$ are the rates of recoils and alpha particle signals per pixel, respectively, $t_{1,2}$- pause times, $\tau_{ER-\alpha}$ recoil-alpha correlation time and $T_{exp}$-the whole experiment duration time. Hence, the total relative loss in the experimental efficiency is given by the formula:

$$\eta = N_{pix} \cdot (\nu_{ER} t_1 + \nu_{ER} \nu_\alpha \tau_{ER-\alpha} t_2).$$

The $N_{pix}$ parameter in the above placed formula is in fact a total effective number of pixels for DSSSD focal plane detector (in our case 48x128 pixels).

Of course, it is very interesting question when is a border between two mentioned sources of a dead time formation. Let us estimate it roughly, namely:

$$\nu_{ER} t_1 \approx \nu_{ER} \nu_\alpha \tau_{ER-\alpha} t_2.$$

Therefore, it should be: $\quad t_1 \leq \nu_\alpha \tau_{ER-\alpha} t_2.$

Note, that in the long term experiment aimed to the synthesis of new super heavy nuclides correlation and pause times are chosen by the experimentalist according to not only rates of recoil/alpha signals, but also according to the predicted theoretically decaying properties of the descendants.

Let us consider additionally the equation system for an optimizing scenario in the form of three questions, namely:

$$\eta(t_1, t_2, \tau_{ER-\alpha}) \leq \mu \ll 1 \;,$$

$$t_1 \leq \nu_\alpha \tau_{ER-\alpha} t_2 \quad ,$$

$$P(\max\{t_1, \tau_{ER-\alpha}\}, \tau_0) \geq 1 - \varepsilon \;.$$

In these formulae are: μ - the acceptable by the experimentalists level of the whole efficiency losses, ε << 1 – small value parameter,  P- probability to detect one decay of nuclide under investigation during $T_{exp}$ time, and $\tau_0$ is the theoretically estimated in advance the life time for the nuclide under investigation. Of course, one extra condition may be as following, especially in the case of very poor statistics:

$$Log\ N_b \leq -N_{min}\ \ .$$

Here, $N_b$ is the expectation parameter value for given multi chain event to be explained by the set of random factors and $N_{min}$ is the accepted by the experimentalist level of statistical significance.

To estimate maximum value of $t_1$ parameter according to equivalent contribution to dead time formation condition, let us consider typical values of $\nu_{ER}$, $\nu_\alpha$, $\tau_{ER-\alpha}$ and $t_2$ as 4.2·10$^{-4}$ c$^{-1}$, 10$^{-4}$ c$^{-1}$, 2 c and 1 min, respectively for about ~ 1pμA beam intensity. In this case the upper reasonable limit for $t_1$ is:

$$t_1 \approx 10^{-4} \cdot 2 \cdot 60 = 12\ ms.$$

The whole loss in the experiment efficiency is estimated as:

$$\eta = N_{pix} \cdot (\nu_{ER}t_1 + \nu_{ER}\nu_\alpha\tau_{ER-\alpha}t_2) = 48 \cdot 128 \cdot 4.2 \cdot 10^{-4}(0.012 + 10^{-4} \cdot 2 \cdot 60) = 0.06.$$

**5. Registered ER energy signal measured with silicon radiation detector**

For successful operation the algorithms reported above, the knowledge about the registered ER energy signal amplitude is strongly required. The multi-parameter events corresponding to production and decays of the super heavy elements usually consist of the time-tagged recoil signal amplitudes and the α-decay signal amplitudes. The amplitudes of the signals associated with one or two fission fragments might be present as well. The pulse amplitudes of ERs and FF (spontaneous fission fragments) are observed with a significant pulse height defect (PHD); nevertheless, they are also of great interest since their presence at the beginning and end of each decay chain makes the identification process complete. A simulation method for modeling of ER spectra obtained from DGFRS is reported in Refs. [4]. ER registered energy spectrum was calculated by a Monte Carlo simulation taking into account neutron evaporation, energy losses in the different media,  energy stragglings, equilibrium charge states distribution width in hydrogen, pulse height defect in PIPS detector, fluctuations of PHD. In [5] a simple empirical equation was obtained as

$$E_{REG} = -2.05 + 0.73 \cdot E_{in} + 0.0015 \cdot E_{in}^2 - (\frac{E_{in}}{40})^3.$$

Here, $E_{in}$ – incoming energy in MeV, $E_{REG}$ - the registered with silicon detector value.

As an example of simulated [6] spectrum in the Fig.9 two ER registered energy spectra are shown. One of them, is indicated by an arrow in the figure ($f_0$), corresponds to no free parameter simulation, whereas for the second one small correction function Err($E_{in}$) is taken into account. This function is obtained experimentally from calibration nuclear reactions. Reaction $^{206}$Pb+$^{48}$Ca→$^{252}$No+2n is one of them. $E_{in}$ is an incoming energy value parameter.

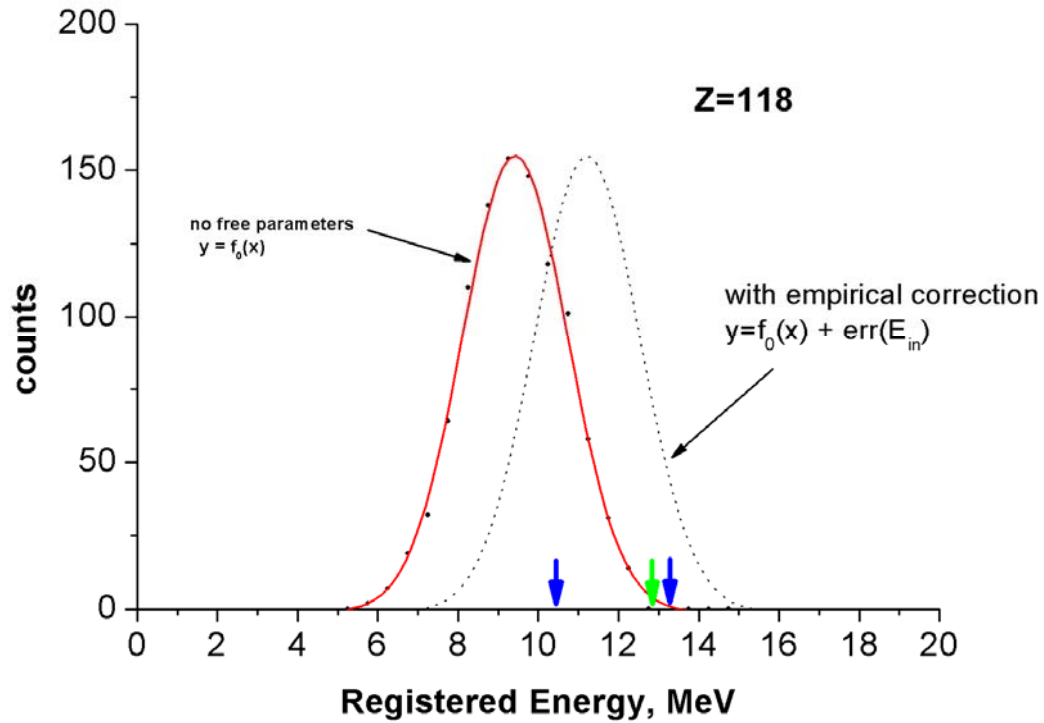

Fig.9 Simulated spectra for Z=118 recoil. Three events detected with silicon radiation detector are shown by arrows [8].

6. **Summary**

Method of "active" correlations was successfully applied in the $^{249}$Bk+$^{48}$Ca→117* experiment when using resistive layer position sensitive PIPS detector. This method was extended to DSSSD detector applications. Real-time method for DSSSD detector to suppress beam associated background products in heavy-ion-induced complete fusion nuclear reactions was implemented and tested in $^{nat}Yb + {}^{48}Ca \rightarrow {}^{217}Th + 3n$ reaction. Neighbor strips edge effects are taken into account in the algorithm development. This method will be applied in the long term experiments aimed to the synthesis of new superheavy isotopes.

As for near-future application of FLNR accelerator facilities with high-intensity (of order 5-10 pµA) heavy-ion beams, the author (Yu. Ts.) does not exclude using correlations of higher order, for instance, ER-α-α, for real-time background suppression. This will be needed if one fails to minimize counting rate in the focal plane detectors by the other, "nonelectronic" methods. The application of DSSSD detectors instead of conventional position-sensitive resistive PIPS detectors can play a large part in the optimization of the technique.

Authors are indebted to DRs A.A.Voinov, V.E.Zhuchko and R.N.Sagaidak for their help. This paper is supported in part by the RFBR grant №13-02-12052.